\documentclass[12pt]{article}
\usepackage{amssymb}
\newcommand{\be}{\begin{eqnarray}}
\newcommand{\ee}{\end{eqnarray}}
\newcommand{\non}{\nonumber}
\newcommand{\n}{\ensuremath{\mathcal{N}}}
\newcommand{\R}{\ensuremath{\mathsf{R}}}
\newcommand{\SSS}{\ensuremath{\mathsf{S}}}
\newcommand{\Y}{\ensuremath{\mathsf{Y}}}

\begin{document}

\begin{titlepage}
\strut\hfill UMTG--230
\vspace{.5in}
\begin{center}

\LARGE Boundary $S$ matrices with $N=2$ supersymmetry\\[1.0in]
\large Rafael I. Nepomechie\\[0.8in]
\large Physics Department, P.O. Box 248046, University of Miami\\[0.2in]  
\large Coral Gables, FL 33124 USA\\

\end{center}

\vspace{.5in}

\begin{abstract}
We propose the exact boundary $S$ matrix for breathers of the $N=2$ 
supersymmetric sine-Gordon model.  We argue that this $S$ matrix has 
three independent parameters, in agreement with a recently-proposed 
action.  We also show, contrary to a previous claim, that the 
``universal'' supersymmetric boundary $S$ matrix commutes with two 
supersymmetry charges.  General $N=2$ supersymmetric boundary 
integrable models are expected to have boundary $S$ matrices with a 
similar structure.
\end{abstract}

\end{titlepage}

\setcounter{footnote}{0}

\section{Introduction}\label{sec:intro}

Much is known about bulk \cite{ZZ} and boundary \cite{GZ} $S$ matrices
of two-dimensional integrable models.  Bulk $S$ matrices $S(\theta)$
for integrable models with $N=1$ or $N=2$ supersymmetry have the
product structure
\be
S(\theta)  = S_{Bose}(\theta)\ S_{SUSY}(\theta)  \,,
\label{bulk}
\ee
where $S_{Bose}(\theta) $ is the $N=0$ bulk $S$ matrix, which is purely
Bosonic; and the ``universal'' $S$ matrix $S_{SUSY}(\theta)$ is an
8-vertex (6-vertex) $R$ matrix satisfying the so-called free-Fermion
condition \cite{FW}, for the case $N=1$ ($N=2$).  See, e.g.,
\cite{SW}-\cite{Ah} for $N=1$, and \cite{BL},\cite{FMVW}-\cite{FI} for
$N=2$, respectively.  One expects that the corresponding boundary $S$
matrices $\SSS(\theta)$ should have a similar structure, \footnote{We
make an effort to distinguish boundary quantities from the
corresponding bulk quantities by using sans serif letters to denote
the former, and Roman letters to denote the latter.}
\be
\SSS(\theta)  = \SSS_{Bose}(\theta)\ \SSS_{SUSY}(\theta)  \,,
\label{boundary}
\ee
where $\SSS_{Bose}(\theta)$ and $\SSS_{SUSY}(\theta)$ satisfy boundary 
Yang-Baxter equations with the corresponding bulk $S$ matrices 
$S_{Bose}(\theta) $ and $S_{SUSY}(\theta)$, respectively.  Indeed, for 
$N=1$, this is precisely what has been found \cite{Ch}-\cite{Ne1}.  
However, for $N=2$, the situation has been less clear: the 
``universal'' $\SSS_{SUSY}(\theta)$ has been claimed \cite{Wa} to 
commute with only one supersymmetry charge, and has therefore been 
rejected in favor of more complicated $S$ matrices with nontrivial 
boundary structure.

We argue here that this $\SSS_{SUSY}(\theta)$ {\it does} in fact
commute with two supersymmetry charges; and hence, the structure
(\ref{boundary}) holds also for the $N=2$ case.  Although this result
is rather general, for definiteness and simplicity we focus here on
the particular case of the first breathers of the $N=2$ supersymmetric
sine-Gordon model.

The outline of this Letter is as follows.  In Section 2 we propose the 
bulk and boundary $S$ matrices for the breathers of the $N=2$ 
sine-Gordon model.  Although the boundary $S$ matrix appears to depend 
on four boundary parameters, we show that there is one relation among 
them.  Hence, there are only three independent boundary parameters, in 
agreement with the recently-proposed action \cite{Ne2}.  In Section 3, 
we verify that the boundary $S$ matrix has $N=2$ supersymmetry. We 
conclude with a brief discussion in Section 4.

\section{The $N=2$ sine-Gordon model}\label{sec:SSG}

Actions for the bulk and boundary $N=2$ sine-Gordon (SG) model have 
been constructed in \cite{KU1} and \cite{Ne2}, respectively. 
Substantial evidence has been given that both the bulk and boundary 
versions of the model are integrable, 
and we now assume that this is indeed the case. The work 
\cite{KU2, FI} suggests that the breathers form two-dimensional 
irreducible representations of the $N=2$ supersymmetry algebra.  
Following \cite{FI}, we denote these one-particle states by 
$u(\theta)$ and $d(\theta)$, where $\theta$ is the 
rapidity.\footnote{As usual, we set $E=m \cosh \theta$, $P= m\sinh 
\theta$, where $m$ is the particle mass.} These states have Fermion 
number ${1\over 2}$ and $-{1\over 2}$, respectively.  For simplicity, 
we restrict our attention to the first (lowest-mass) breathers.
In view of (\ref{bulk}), a natural conjecture for the bulk 
(two-particle) $S$ matrix $S^{N=2}_{SG\ b}(\theta, \beta^{N=2})$ is
(see also \cite{KU2, FI})
\be
S^{N=2}_{SG\ b}(\theta, \beta^{N=2}) = 
S^{N=0}_{SG\ b}(\theta, \beta^{N=0})\ S_{SUSY}(\theta) \,,
\label{bulkSmatrix}
\ee
where $\theta$ is the difference in rapidity of the two particles, and 
the scalar factor $S^{N=0}_{SG\ b}(\theta, \beta^{N=0})$ is the $N=0$ 
sine-Gordon breather $S$ matrix \cite{AK, ZZ}
\be
S^{N=0}_{SG\ b}(\theta, \beta^{N=0}) 
= {\sinh \theta + i \sin {\gamma\over 8}\over
\sinh \theta - i \sin {\gamma\over 8}} \,, 
\ee
where $\gamma = \beta_{N=0}^{2}/ (1-(\beta_{N=0}^{2}/ 8\pi))$.  The 
universal $N=2$ bulk $S$ matrix $S_{SUSY}(\theta)$ is the sine-Gordon 
soliton $S$ matrix \cite{ZZ} with $\beta^{2}= {16\pi\over 3}$,
\be
S_{SUSY}(\theta) = S^{N=0}_{SG\ s}(\theta, 
\beta^{2}= {16\pi\over 3}) = Y(\theta) R(\theta) \,,
\ee
where $R(\theta)$ is the $4 \times 4$ matrix 
\be
R(\theta) = \left( \begin{array}{cccc}
	a             &0         &0           &0             \\
        0             &b         &c           &0            \\
	0             &c         &b           &0            \\
	0             &0         &0           &a 
\end{array} \right) \,, 
\label{bulkRmatrix}
\ee 
with matrix elements
\be
a  = i\cosh  {\theta\over 2}
\,, \qquad
b   = \sinh  {\theta\over 2}
\,, \qquad
c=i \,.
\label{Rmatrixelements}
\ee
The scalar factor $Y(\theta)$ is given by
\be
Y(\theta) = {1\over i \cosh{\theta\over 2}}
\exp \left( {i\over 2}\int_{0}^{\infty} {dt\over t}\
{\sin (t \theta) \over \cosh^{2}{\pi t\over 2}} \right) 
\,.
\label{bulkYfactor}
\ee
Finally, $\beta^{N=2}$ is the dimensionless bulk coupling constant 
appearing in the action, which is related to $\beta^{N=0}$ 
by \cite{KU2, FI}
\be
\beta_{N=2}^{2}=\gamma = 
{\beta_{N=0}^{2}\over 1-{\beta_{N=0}^{2}\over 8\pi}} \,.
\ee 

For the corresponding breather boundary $S$ matrix 
$\SSS^{N=2}_{SG\ b}(\theta, \beta^{N=2})$, we propose
\be
\SSS^{N=2}_{SG\ b}(\theta, \beta^{N=2})= 
\SSS^{N=0}_{SG\ b}(\theta, \beta^{N=0}\,;  
\tilde\eta \,, \tilde\vartheta)\ 
\SSS_{SUSY}(\theta \,;  \eta \,, \vartheta ) \,,
\label{boundarySmatrix}
\ee
where the scalar factor 
$\SSS^{N=0}_{SG\ b}(\theta, \beta^{N=0}\,;  
\tilde\eta \,, \tilde\vartheta)$
is the $N=0$ sine-Gordon breather boundary $S$ matrix \cite{Gh}
\be
\SSS^{N=0}_{SG\ b}(\theta, \beta^{N=0}\,;  
\tilde\eta \,, \tilde\vartheta)
&=&  {\cosh ({\theta\over 2} + {i\gamma\over 32})
\cosh ({\theta\over 2} - {i \pi\over 4} - {i\gamma\over 32})
\sinh({\theta\over 2} + {i \pi\over 4})\over
\cosh ({\theta\over 2} - {i\gamma\over 32})
\cosh ({\theta\over 2} + {i \pi\over 4} + {i\gamma\over 32})
\sinh({\theta\over 2} - {i \pi\over 4})} \non \\
&\times&
{\Big(\cos ({\gamma \tilde\eta\over 8 \pi}) + i \sinh \theta \Big)
 \Big(\cosh ({\gamma \tilde\vartheta\over 8 \pi}) 
 + i \sinh \theta \Big)\over 
 \Big(\cos ({\gamma \tilde\eta\over 8 \pi}) - i \sinh \theta \Big)
 \Big(\cosh ({\gamma \tilde\vartheta\over 8 \pi}) 
 - i \sinh \theta \Big)}
\,. 
\label{ghoshal}
\ee
Moreover, $\SSS_{SUSY}(\theta \,; \eta \,, \vartheta )$ is the 
sine-Gordon soliton boundary $S$ matrix \cite{GZ} with $\beta^{2}= 
{16\pi\over 3}$,
\be
\SSS_{SUSY}(\theta \,; \eta \,, \vartheta) =
\SSS^{N=0}_{SG\ s}(\theta, \beta^{2}= {16\pi\over 3}\,; 
\eta \,, \vartheta)
= \Y(\theta \,; \eta \,, \vartheta)\ 
\R(\theta \,; \eta \,, \vartheta) \,,
\label{boundarySUSY}
\ee 
where $\R(\theta \,; \eta \,, \vartheta)$ is the $2 \times 2$ matrix
\be
\R(\theta \,;  \eta \,, \vartheta) = \left( \begin{array}{cc}
{\cal A}_{+} &{\cal B} \\
{\cal B} & {\cal A}_{-} 
\end{array} \right) \,, 
\label{boundaryRmatrix}
\ee 
with matrix elements
\be
{\cal A}_{\pm} = \cos(\xi \mp {i\theta\over 2}) \,, \qquad 
{\cal B} = -{i k\over 2}\sinh \theta  \,.
\label{boundelem1}
\ee
The pair of boundary parameters $(\xi \,, k)$ is related to 
$(\eta \,, \vartheta)$ by 
\be
\cos \eta \cosh \vartheta = -{1\over k}\cos \xi \,, \qquad
\cos^{2} \eta + \cosh^{2} \vartheta = 1 + {1\over k^{2}} \,.
\ee
An explicit expression for the scalar factor $\Y(\theta \,; \eta \,, 
\vartheta)$, which we shall not need here, can be obtained from \cite{GZ}.
As usual \cite{ZZ, GZ}, the $S$ matrices (\ref{bulkSmatrix}), 
(\ref{boundarySmatrix}) may contain additional CDD-like factors.

The boundary $S$ matrix (\ref{boundarySmatrix}) apparently depends on 
four boundary parameters: $\tilde\eta \,, \tilde\vartheta \,, \eta \,, 
\vartheta$.  However, they are not all independent. For example, let 
us suppose that $\tilde\eta$ lies in the range
\be
{4\pi^{2}\over \gamma} < \tilde\eta < {8\pi^{2}\over \gamma} \,.
\label{range}
\ee
The scalar factor (\ref{ghoshal}) has a pole
at $\theta = i v$ with $v={\gamma\tilde\eta\over 8\pi} - {\pi\over 
2}$, which then lies in the physical strip and corresponds to a 
boundary bound state. We recall \cite{GZ} the following general 
constraint: near a pole $i v^{\alpha}_{0 a}$ of the boundary $S$ 
matrix associated with the excited boundary state $|\alpha 
\rangle_{B}$ (which can be interpreted as a boundary bound state of 
particle $A_{a}$ with the boundary ground state $|0\rangle_{B}$), the 
boundary $S$ matrix must have the form
\be
\SSS_{a}^{b}(\theta) \simeq {i\over 2}
{g^{\alpha}_{a 0} g^{b 0}_{\alpha}\over \theta - i v^{\alpha}_{0 a}} \,,
\ee 
where $g^{\alpha}_{a 0}$ are boundary-particle couplings.  
For the boundary $S$ matrix (\ref{boundarySmatrix}), this 
implies \footnote{There is a similar relation for the $N=1$ case 
\cite{AN2, Ne1}.}
\be
\left( {\cal A}_{+}{\cal A}_{-} - {\cal B}^{2} \right)
\Big\vert_{\theta = iv} = 0 \,.
\label{constraint}
\ee 
Remarkably, we find (after some computation) that this constraint 
reduces to simply $v= 2\eta \pm \pi$. \footnote{There is another 
solution $v= 2i\vartheta \pm \pi$, which we discard for $\vartheta$ 
real.} It follows that 
\be
\eta = {\gamma \tilde\eta\over 16\pi} + {\pi\over 4}\,,
\ee 
and so there are indeed only three independent boundary 
parameters. (For values of $\tilde\eta$ outside the range 
(\ref{range}), one must presumably consider $S$ matrices for the 
higher breathers and/or solitons.)
The fact that the boundary action \cite{Ne2} has the same 
number of parameters lends support to the proposed expression 
(\ref{boundarySmatrix}) for the boundary $S$ matrix.  In the next 
section, we provide further support by demonstrating that this $S$ 
matrix has $N=2$ supersymmetry.

\section{$N=2$ supersymmetry of the $S$ matrix}\label{sec:symmetry}

Following \cite{FI}, we assume that the supersymmetry charges 
$Q^{\pm}$, ${\overline Q}^{\pm}$ act on the one-particle states as 
follows:
\be
Q^{-} | u(\theta) \rangle &=& \sqrt{2m}e^{\theta\over 2} | d(\theta) 
\rangle \,, \qquad 
{\overline Q}^{+} | u(\theta) \rangle = \sqrt{2m}e^{-{\theta\over 2}}
| d(\theta) \rangle \,, \non \\
Q^{+} | d(\theta) \rangle &=& \sqrt{2m}e^{\theta\over 2} | u(\theta) 
\rangle \,, \qquad 
{\overline Q}^{-} |d(\theta) \rangle = \sqrt{2m}e^{-{\theta\over 2}} 
| u(\theta) \rangle \,, 
\label{actiononstates}
\ee
and otherwise annihilate the states. The action on multi-particle states 
is specified further by the coproduct
\be
\Delta(Q^{\pm}) &=& Q^{\pm} \otimes 1 + e^{\pm i \pi F} \otimes 
Q^{\pm} \,, \non  \\
\Delta({\overline Q}^{\pm}) &=& {\overline Q}^{\pm} \otimes 1 
+ e^{\mp i \pi F} \otimes {\overline Q}^{\pm} \,,
\label{coproduct}
\ee
where $F$ is the Fermion number operator,
\be
F | u(\theta) \rangle = {1\over 2} | u(\theta) \rangle \,, \qquad 
F | d(\theta) \rangle = -{1\over 2} | d(\theta) \rangle \,.
\label{fermionnumber}
\ee 

One can verify that (\ref{actiononstates})-(\ref{fermionnumber}) 
indeed provide a representation of the $N=2$ supersymmetry 
algebra \cite{WO}
\be
\left\{ Q^{+} \,, Q^{-} \right\} &=& 2(H+P) \,, \qquad 
\left\{ {\overline Q}^{+} \,, {\overline Q}^{-} \right\} = 2(H-P) \,,
\non  \\ 
\left\{ Q^{+} \,, {\overline Q}^{+} \right\} &=& 2m \n \,, \qquad 
\qquad 
\left\{ Q^{-} \,, {\overline Q}^{-} \right\} = 2m \n \,,
\non  \\ 
Q^{\pm 2} = {\overline Q}^{\pm 2} &=& 
\left\{ Q^{\pm} \,, {\overline Q}^{\mp} \right\} = 0 \,, \qquad
\left\{ Q^{\pm} \,, F \right\} =
\left\{ {\overline Q}^{\mp} \,, F  \right\} = 0 \,,
\ee
where $H$, $P$ and $\n$ are the Hamiltonian, momentum and number
operators, respectively.  One can also verify that the bulk $S$ matrix
(\ref{bulkSmatrix}) commutes with the supersymmetry charges $Q^{\pm}$,
${\overline Q}^{\pm}$, as well as with $F$.

We can now address the important question: how much supersymmetry does 
the boundary $S$ matrix (\ref{boundarySUSY}) have?  Evidently, this 
$S$ matrix does not commute with any of the supersymmetry charges or 
with the Fermion number operator.  But it is not difficult to prove 
that there are two linear combinations of these operators with which 
the boundary $S$ matrix {\it does} commute:
\be
\widehat Q^{+} &=& Q^{+} + {\overline Q}^{+} + \kappa^{+} F \,, \non \\
\widehat Q^{-} &=& Q^{-} + {\overline Q}^{-} + \kappa^{-} F \,,
\label{nicecombos}
\ee
where $\kappa^{\pm} = \pm i 2 \sqrt{2m} e^{\pm i \xi}/k$.  This means
that the boundary $S$ matrix has $N=2$ supersymmetry.  The terms in
(\ref{nicecombos}) proportional to $F$ correspond to local
Fermionic boundary terms \cite{Ne2}.  Similar Fermionic boundary terms
also appear in the $N=1$ case \cite{AN2, Ne1}.  It was the failure to
consider such terms in \cite{Wa} that led to the erroneous conclusion
that the $S$ matrix has only $N=1$ supersymmetry.

We remark that $\widehat Q^{\pm}$ generate the subalgebra
\be
\widehat Q^{\pm 2} = 2 m \n + \kappa^{\pm 2}F^{2} \,, \qquad
\left\{ \widehat Q^{+} \,, \widehat Q^{-} \right\} 
= 4 H + 2\kappa^{+}\kappa^{-}F^{2} \,.
\ee
We also note that combinations of the form 
$Q^{\pm} + a_{\pm} {\overline Q}^{\mp} + b_{\pm} F$ generally do not 
commute with the boundary $S$ matrix.

\section{Discussion}\label{sec:discuss}

We have argued that the boundary $S$ matrix of the $N=2$ sine-Gordon
model has the structure (\ref{boundary}), with the universal $S$ matrix
$\SSS_{SUSY}(\theta)$ given by (\ref{boundarySUSY}).  On the basis of
established corresponding bulk results (\ref{bulk}), we expect that
this structure is characteristic of all boundary integrable models
with $N=2$ supersymmetry.

An important outstanding problem is to find the precise relation 
between the parameters of the boundary $S$ matrix and those of the 
boundary action.  This problem has already been addressed for the 
$N=0$ case \cite{AlZ, Co}.

Whereas for the $N=1$ case there is only one boundary parameter in the 
universal $S$ matrix $\SSS_{SUSY}$, for the $N=2$ case there are two.
Hence, $N=2$ models generally should manifest much richer boundary 
phenomena. It should be particularly interesting to explore the 
consequences of this for open string theory.

\section*{Acknowledgments}

Discussions with L. Mezincescu and A.B. Zamolodchikov in 1995 on 
related problems are gratefully acknowledged.
This work was supported in part by the National Science Foundation
under Grant PHY-9870101.


\begin{thebibliography}{99}

\bibitem{ZZ}
A.B. Zamolodchikov and Al.B. Zamolodchikov, Ann. Phys. {\it 120} (1979) 253;
A.B. Zamolodchikov, Sov. Sci. Rev. {\it A2} (1980) 1.

\bibitem{GZ}
S. Ghoshal and A.B. Zamolodchikov, Int. J. Mod. Phys. {\it A9} (1994)
3841.
   
\bibitem{FW}
C. Fan and F.Y. Wu, Phys. Rev. {\it B2} (1970) 723.

\bibitem{SW}
R. Shankar and E. Witten, Phys. Rev. {\it D17} (1978) 2134.

\bibitem{Za1} 
A.B. Zamolodchikov, ``Fractional-spin integrals of motion in perturbed
conformal field theory,'' in {\it Fields, Strings and Quantum Gravity}, eds.
H. Guo, Z. Qiu and H. Tye, (Gordon and Breach, 1989).

\bibitem{Sc}
K. Schoutens, Nucl.Phys. {\it B344} (1990) 665.

\bibitem{BL}
D. Bernard and A. LeClair, Phys. Lett. {\it 247B} (1990) 309.

\bibitem{ABL}
C. Ahn, D. Bernard and A. LeClair, Nucl. Phys. {\it B346} (1990) 409.

\bibitem{Ah}
C. Ahn, Nucl. Phys. {\it B354} (1991) 57; {\it B422} (1994) 449.

\bibitem{FMVW}
P. Fendley, S.D. Mathur, C. Vafa and N.P. Warner, Phys. Lett.
{\it 243B} (1990) 257.

\bibitem{FLMW}
P. Fendley, W. Lerche, S.D. Mathur and N.P. Warner, Nucl. Phys.
{\it B348} (1991) 66.

\bibitem{MW}
P. Mathieu and M.A. Walton, Phys. Lett. {\it 254B} (1991) 106.

\bibitem{KU1} 
K. Kobayashi and T. Uematsu, Phys. Lett. {\it B264} (1991) 107.

\bibitem{KU2} 
K. Kobayashi and T. Uematsu, Phys. Lett. {\it B275} (1992) 361.

\bibitem{FI}
P. Fendley and K. Intriligator, Nucl. Phys. {\it B372} (1992) 533;
{\it B380} (1992) 265.

\bibitem{Ch}
L. Chim, Int. J. Mod. Phys. {\it A11} (1996) 4491.

\bibitem{MS}
M. Moriconi and K. Schoutens, Nucl. Phys. {\it B487} (1997) 756.

\bibitem{AR}
C. Ahn and C. Rim, J. Phys. {\it A32} (1999) 2509.

\bibitem{AN1}
C. Ahn and R.I. Nepomechie, Nucl. Phys. {\it B586} (2000) 611.

\bibitem{AN2}
C. Ahn and R.I. Nepomechie, Nucl. Phys. {\it B594} (2001) 660.

\bibitem{Ne1}
R.I. Nepomechie,  Phys. Lett. {\it B509} (2001) 183.

\bibitem{Wa}
N.P. Warner, Nucl. Phys. {\it B450} (1995) 663.

\bibitem{Ne2}
R.I. Nepomechie, ``The boundary $N=2$ supersymmetric sine-Gordon 
model,'' {\tt hep-th/0106207}.

\bibitem{AK}
I.Ya. Aref'eva and V.E. Korepin, JETP Lett. {\it 20} (1974) 312.

\bibitem{Gh}
S. Ghoshal, Int. J. Mod. Phys. {\it A9} (1994) 4801.

\bibitem{WO}
E. Witten and D. Olive, Phys. Lett. {\it 78B} (1978) 97.

\bibitem{AlZ}
Al.B. Zamolodchikov, unpublished work reported at the 1999 Bologna  
CFT workshop.

\bibitem{Co}
A. Chenaghlou and E. Corrigan, Int. J. Mod. Phys. {\it A15} (2000) 4417;
E. Corrigan and A. Taormina, J. Phys. {\it A33} (2000) 8739.

\end{thebibliography}
\end{document}